\documentclass{aa}  

\usepackage{graphicx}
\usepackage{color}
\usepackage{txfonts}
\begin{document}

   \title{Gas dynamics around a Jupiter-mass planet\\  I. Influence of protoplanetary disk properties}
   \subtitle{}

   \author{E. Lega\inst{1}, M. Benisty\inst{1,2}, A. Cridland\inst{1,2,3}, A. Morbidelli\inst{4,1}, 
    M. Schulik\inst{5,6} and M. Lambrechts\inst{7}}

   \institute{
   Université Côte d'Azur, Observatoire de la Côte d'Azur, CNRS, Laboratoire Lagrange, France\label{inst1}
   \and
   Univ. Grenoble Alpes, CNRS, IPAG, 38000 Grenoble, France\label{inst2}
   \and 
   Universit\"ats-Sternwarte M\"unchen, Ludwig-Maximilians-Universit\"at, Scheinerstr. 1, 81679 M\"unchen, Deutschland
   \label{inst3} 
   \and
   Collège de France, 11 Pl. Berthelot, 75005 Paris, France \label{inst4}
   \and
   Imperial College London, Kensington Lane, London W1J 0BQ, UK
\label{inst5}
\and
Department of Earth, Planetary, and Space Sciences, University of California, LosAngeles, CA90095,USA \label{inst6}
\and
Center for Star and Planet Formation, Globe Institute, University of Copenhagen, {{\O}}sterVoldgade 5-7, 1350 Copenhagen, Denmark
\label{inst7}
   }

   \date{Received S; accepted }

\abstract
{Giant planets grow and acquire their gas envelope during the disk phase.  At the time of the discovery of giant planets in their host disk, it is important to understand the interplay between the host disk and the envelope and circum-planetary disk properties of the planet.}
{Our aim is to investigate the dynamical and physical structure of
the gas in the vicinity of a Jupiter-mass planet and study how protoplanetary disk properties, such as disk mass and viscosity, determine the planetary system as well as the accretion rate inside the planet's Hill sphere.}
{We ran global 3D simulations with the grid-based code {\it fargOCA,} using a fully radiative equation of state and a dust-to-gas ratio of 0.01. We built a consistent disk structure starting from vertical thermal equilibrium obtained by including stellar irradiation. We then let a gap open with a sequence of phases, whereby we deepened the potential and increased the resolution in the planet's neighbourhood. We explored three models. The nominal one features a disk with surface density,  $\Sigma$, corresponding to the minimum mass solar nebula at the planet's location (5.2 au), characterised by an $\alpha$ viscosity value of $4\,10^{-3}$ at the planet's location. The second model has a surface density that is ten times smaller than the nominal one and the same viscosity. In the third model, we also reduced  the viscosity value by a factor of 10.}
{During gap formation, giant planets accrete gas inside the Hill sphere   from the local reservoir. Gas is heated by compression and cools according to opacity, density, and temperature values. This process determine the thermal energy budget inside the Hill sphere.  In the analysis of our disks, we find  that the gas flowing into the Hill sphere is  approximately scaled as the product  $\Sigma \nu $, as expected from viscous transport.  The  accretion  rate of the planetary system (envelope plus circum-planetary disk) is instead scaled   as$~\sqrt\Sigma \nu $, with its efficiency depending on the thermal energy budget inside the Hill sphere.}
 {Previous studies have shown that pressure-supported or rotationally supported structures are formed around giant planets, depending on the equation of state (EoS) or on the opacity; namely, on the dust content within the Hill sphere. In the case of a fully radiative EoS  and a constant dust to gas ratio of 0.01, we find that low-mass and low-viscosity circum-stellar disks favour  the formation of a  rotationally supported  circum-planetary disk. Gas accretion leading to the doubling time of the planetary system of  $>10^{5}$ years has only been found in the case of a low-viscosity disk.}


\titlerunning{Gas dynamics around a Jupiter-mass planet}
\authorrunning{Lega et al.} 

\keywords{Hydro-dynamical simulations, planet-disk interactions, circum-planetary disks, giant planets}

\maketitle

\section{Introduction}
The discovery of  PDS70b and  PDS70c, the first protoplanets directly imaged in their host disk  \citep{Keppler2018,Haffert2019}, provides a unique opportunity to gain new insights in the processes of planet formation and evolution and how these giant planets acquired their atmospheric composition \citep{Cridland2023}. PDS70b and c are giant planets located in a large dust-depleted cavity, at 21 and 34\,au, respectively. Their properties remain unclear but all studies are in agreement with  the assumption of high masses \citep[3-12\,M$_{\rm{Jup}}$][]{Mueller2018,Haffert2019,Stolker2020,wang2020}. 
Interestingly, both planets are aptly detected in the H$\alpha$ line and PDS70c is also detected in the UV continuum \citep[e.g.][]{Haffert2019,Hashimoto2020, Zhou2021}, indicating ongoing accretion of gas onto the planets and possibly accretion shocks \citep{Aoyama2018}. Spectro-photometric studies  show infrared spectral energy distributions, which are notably red and flat, indicating the existence of dust located within the planetary atmosphere or in a surrounding circum-planetary disk \citep{christiaens2019,wang2021}. The presence of small grains around both planets (and in the inner disk within the cavity) is also suggested by hydro-dynamical simulations by \citet{bae2019}. Interestingly, evidence for dust grains co-located with PDS70c and b was found with ALMA sub-millimeter continuum observations \citep{Isella2019, Benisty2021}. While the emission around PDS70c, closer to the outer protoplanetary disk, is compact and unresolved (at 2\,au resolution), the emission around PDS70b, which is located further-in within the dust-depleted cavity, is faint and diffuse, with a morphology that remains unclear. It is thought that PDS70b is starved of dust grains from the outer disk, as they are filtered out by PDS70c (Pinilla et al. in prep). While both planets seem to be surrounded by dust, the different observational properties of the emissions around PDS70b and c calls for a better understanding of the structure of the material around the planets and, in general, of the conditions for circum-planetary disks to form. 


In this context, numerical simulations can guide our understanding and address major questions such as the formation of a circum-planetary disk (CPD, hereafter), its structure and accretion properties. 
From the numerical perspective studying the gas dynamics around a giant planet is challenging. As giant planets accrete from the protoplanetary disk and part of the gas  enters in the Hill sphere, numerical models require us to resolve both the full disk behaviour and the dynamics in the planet's Hill sphere. This goal is achieved by using  nested meshes on grid based codes either on the full disk or with a local approach, namely, shearing sheet approximation \citep{2003ApJ...586..540D,2006A&A...445..747K,2008ApJ...685.1220M,2012ApJ...747...47T,Szulagyietal2014,Szulagyietal2016}, non-regular grid geometry in the planet's Hill sphere \citep{LL17,2019ApJ...887..152F,LLNCM19,2019A&A...632A.118S,Schuliketal2020}, or using SPH codes \citep{2009MNRAS.397..657A}.  

The first studies have been done in 2D models \citep{2002A&A...385..647D} 
and extended to three dimensions as soon as computing capabilities allowed for it. The reason for including  the vertical dimension  for giant planets is that their Hill radius is comparable to disk scale height
and, in fact, vertical motion has been shown to have an
important role for a planet's accretion (see for example \cite{2003MNRAS.341..213B,2006A&A...445..747K,2012ApJ...747...47T,Szulagyietal2014,MorbyMeridional2014}).
Comparing the numerical results from the literature is not a simple task, as different codes are used and different physical models are involved. Some of the studies consider  the isothermal approximation \citep{2008ApJ...685.1220M,2010MNRAS.405.1227M,2012ApJ...747...47T,Szulagyietal2014,2019ApJ...887..152F,2019MNRAS.488.2365B}, while others adopt a more complex equation of state  (EoS hereafter) going from adiabatic \citep{2010MNRAS.405.1227M,2019ApJ...887..152F} to fully radiative EoS on inviscid  or  viscous disks \citep{2003ApJ...586..540D,2006A&A...445..747K,2009MNRAS.397..657A,Szulagyietal2016,LL17,LLNCM19,2019A&A...632A.118S,Schuliketal2020,2023ApJ...952...89M}; in this case, the   opacity depends on physical disk properties such as the density and the temperature, or is simply considered as constant  \citep{Schuliketal2020}. 


A common result of isothermal models is the formation of a CPD around giant planets. All models agree that the gas that reaches the planetary system  (planet plus circum-planetary disk) mainly comes from polar regions and  flows down towards the planet  with a spiral pattern. A highly debated question among the models is whether gas is also accreted from the equatorial direction. In the case of inviscid disks, some authors   have shown that gas reaching the planetary system from disk midplane may lose angular momentum due to the torque exerted by the star through the spiral wave \citep{Szulagyietal2014}, whilst others \citep{2006A&A...445..747K,2008ApJ...685.1220M,2012ApJ...747...47T} have not found any midplane accretion. However, even in the case of midplane accretion, \citet{Szulagyietal2014} found that only $10\%$ of the accreted gas comes from disk midplane, while the remaining accretion was due to polar flow. 

In contrast, when considering fully radiative EoS, previous studies are not in agreement: some authors find fully rotationally supported CPD \citep{2009MNRAS.397..657A,2023ApJ...952...89M}, whilst others report a slow rotation and a pressure supported planetary envelope \citep{2006A&A...445..747K,Szulagyietal2016,LLNCM19,2024arXiv240214638K}. Using the adiabatic approximation,  \cite{2019ApJ...887..152F} showed  that a  spherical pressure supported adiabatic envelope is found instead of a CPD. Therefore, the isothermal and the adiabatic approximations give the  two opposite situations of  an extremely efficient and inefficient cooling in the planet's environment and   we can expect to find either a pressure-supported envelope or a CPD when considering the more realistic fully radiative EoS  model, depending on the planet's environment.
\par
Concerning the vertical inflowing   gas, it is travelling nearly in free fall at supersonic speed in the cases where a rotationally supported CPD is found  \citep{2019ApJ...887..152F,2003ApJ...586..540D,2023ApJ...952...89M} and subsonic speeds are reported for pressure supported adiabatic envelopes \citep{2019ApJ...887..152F,2019MNRAS.488.2365B}. 
The  inflowing gas hits the envelope's surface or the CPD and if the shocks occurs at a velocity above a given critical value,  the emission of hydrogen  $H_{\alpha}$ lines is expected \citep{2003ApJ...586..540D,2023ApJ...952...89M}. \par The outcome of fully radiative models has been shown to depend on the temperature at the planet's position
\citep{Szulagyietal2016}, which  can itself be regulated by  the opacity \citep{2009MNRAS.397..657A,2019A&A...632A.118S,Schuliketal2020}.
More precisely, keeping the gas bound to the planet  cold or  decreasing the dust-to-gas ratio to have a low opacity in the local environment enhances the rotational support with respect to the pressure support. The gas accretion rate onto the planet seems independent on the rotational support of CPD: typical accretion rates  for a Jupiter-mass planet take values of  the order of $2-6 \times 10^{-5}$ Jupiter masses per year \citep{2003MNRAS.341..213B,2006A&A...445..747K,2009MNRAS.397..657A,2010MNRAS.405.1227M,LLNCM19,Schuliketal2020}; with a larger value of $10^{-4}$ Jupiter masses per year in the case of a low opacity disk found in \citep{Schuliketal2020}. Even the lower value of $2 \,10^{-5}$ Jupiter masses per year gives a doubling mass time an order of magnitude shorter than typical 
disk lifetime, enforcing the idea that Jupiter-mass planets emerged late in disk lifetime.  A decline of mass accretion rate is expected for planets embedded in wide gaps \citep{2010MNRAS.405.1227M}.
This could be the case of giant planets embedded in low-viscosity disks. \par
Recently, \cite{2023A&A...670A.113N}  measured accretion rates on Jupiter-mass planets by using a simple hydrodynamical model with  star accretion on surface layers induced through angular momentum loss. These authors found accretion rates in the interval  $[10^{-7},4\times 10^{-6}]$ Jupiter masses per year. This result does not require giant planets to form later on.
Moreover, the study of low-viscosity disks has garnered interest  in recent years both from observational  constraints \citep{pinte_dust_2016,villenave_highly_2022} and from theories revisiting the generation of turbulence in disks  via the magneto-rotational instability \citep{turner_transport_2014}.  Recent magnetohydrodynamical models \citep{2017ApJ...845...75B,2017A&A...600A..75B} that take into account non-ideal effects  have shown that in disk with a low bulk viscosity ($\alpha \sim 10^{-5}$), gas can be transported to the star by thin accretion layers  located at the disk surface at rate compatible with the observed star-accretion rates. \par
The aim of this paper to describe the dynamics and  physical structure of the gas in the immediate vicinity of a giant planet and investigate the role of key properties of the protoplanetary disk, such as the
mass or the viscosity in determining the  outcome of the CPD and the planet's accretion rate. 
We follow a self-consistent approach that goes from the study of the disk structure, whilst also considering the formation of a gap in the protoplanetary disk. This paper is the first of a series of (at least) two papers, with paper II focusing on the chemical delivery of volatile to the planet (\cite{Cridland2024}, submitted). 
The paper is organised as follows. In Sect. \ref{Sec:model}, we present the physics and the numerical models.  We also compare global properties of the three disk models with an embedded  Jupiter-mass planet. In Sect. \ref{Sec:Dynamics}, we provide  details of the gas dynamics focusing on the Hill sphere. Our conclusions are given in Sect.\ref{Sec:Conclusion}.

\section{Models}
\label{Sec:model}
\subsection{Physical model}
In this work, the protoplanetary disk (PPD hereafter) is treated as a non-self-gravitating gas whose motion is described by Navier-Stokes equations. We used the grid-based  \textit{fargOCA} code \citep{2014MNRAS.440..683L} with a specific non-uniform grid setting introduced in \citep{LL17}.
We used spherical coordinates $(r,\varphi,\theta),$ 
where $r$ is the radial distance from the star (which is at the origin of the coordinate system), $\varphi$ is the azimuthal coordinate measured from the $x$-axis, and $\theta$ is the 
polar angle measured from the $z$-axis (the colatitude). The midplane of the disk is located at the equator,
$\theta = \frac {\pi} {2}$. 
We worked in a coordinate system that rotates with the angular velocity of a planet of a mass, $m_p$:
$$\Omega_p = \sqrt {\frac{G(M_{\star}+m_p)}{{r_p}^3}}, $$
where $M_{\star}$ is the mass of the central star, $G$ is the gravitational constant,
and $r_p$ is the semi-major axis of  a planet  assumed to be on a circular orbit.
We consider a planet orbiting a Solar-mass star and, therefore,  we indicate the
mass of the star as $M_{\sun}$ in the following.

\par
The dynamics of the gas is provided by the integration of the  Navier-Stokes equations composed by a continuity equation and a set of three equations for the momenta \citep{2014MNRAS.440..683L}.
To the Navier-Stokes system of equations, we  added an equation for the energy density of the thermal radiation, $E_r$,  and that of the internal energy, $e=\rho c_v T$, where $\rho$
and $T$ are the gas volume density and the gas temperature, whilst 
$c_v$ is the specific heat at constant volume.
We followed the evolution of both quantities using the so-called two temperature approach \citep{Commercon2011}:
\begin{equation}
\label{eq:Edot2Temp}
\left\lbrace \begin{array}{lll}
  \frac{\partial E_{\rm r}}{\partial t} - \nabla \cdot  \vec F & = &
    \rho \kappa_p(a_rT^4 - cE_{\rm r}), \\
  \frac{\partial e}{\partial t} +  \nabla \cdot(e\vec v) &= 
&
  -P \nabla \cdot \vec v -\rho \kappa_p(a_rT^4 - cE_{\rm r})  + Q^+,
\end{array} \right.
\end{equation}

where  $\vec F =   \frac{c\lambda}{\rho \kappa_r} \nabla E_{\rm r}$ is the radiation flux vector calculated in the flux limited diffusion approximation \citep[FLD,][]{LevermorePomraning81} with $\lambda$ as the flux limiter \citep{KBK09}.
We indicate with $\kappa_p$ and  $\kappa_r$   respectively the Planck  and the Rosseland mean opacity, $\sigma$ is the Stefan-Boltzmann constant, $c$ is the speed of light.
In this paper we consider $\kappa_p = \kappa_r$ (see \cite{Bitschetal2013}) and use the opacity law of \cite{1994ApJ...427..987B}.
 \par We considered an ideal gas of pressure $P$  with equation of state: 
 $P = (\gamma-1)e$ for a  gas with  adiabatic index of $\gamma=1.4$.
 The terms $P \nabla \cdot \vec v $  and $Q^+$  are  the compressional heating and the viscous heating, respectively \citep{MihalasMihalas84}.
 We did not include the heating from the central star. This choice allows us to  reduce the computational cost with no significant impact on the gas dynamics in the planet vicinity \citep[see][]{Legaetal2015}.
 However, the heating from the central star has an impact on the thermal properties of the disk. In  Appendix \ref{sec:AppendixA}, we describe the procedure used to find the disk surface temperature that corresponds to the disk aspect ratio in the  vicinity  if the planet relative to that obtained for a disk heated by the central star.

\subsection{Disk models}
 We modelled a full disk annulus with an azimuthal domain in $[-\pi,\pi]$ since we are interested in the analysis of the exchange of gas between the planet's Hill sphere and the protoplanetary disk.
We are also required to trace the gas dynamics on small scales well inside the Hill sphere. For this purpose, we use a non-uniform grid geometry with a grid suitably designed to have small and nearly uniform cells around the planet and large grid-cells further out. This procedure was introduced by \citet{Fung15} and tested for the code \textit{fargOCA} by \citep{LL17,LLNCM19}.
We remark that our small scales inside the Hill sphere correspond to about 10 Jupiter radii (considering Jupiter's current radius); therefore, we cannot study the phase of growth of Jupiter nor define a boundary between the planet and the surrounding gas. In this paper, we consider three simulations  of a disk with constant viscosity and surface density defined as:
$\Sigma  = \Sigma_0 (r/r_p)^{-1/2}$:

(1) {\bf The `nominal' case}, hereafter referred to as {\bf N}, for which we consider $(\Sigma_0 / M_{\sun}r_p^{-2}) =  6.67 \times 10^{-4}$,  corresponding to about $210 g/cm^2$ at the planet's location.  For this simulation the viscosity is $\nu / r^2_p\Omega_p = 10^{-5} $ in code units (the value $ 10^{-5} $ in code units corresponds to  $10^{11}\rm{m^2/s}$ or $10^{15} \rm{cm^2/s}$ and to a value of $\alpha = 4\, 10^{-3}$ at the planet's position of 5.2 au). These disk parameters are the same as those reported in  \citet{Szulagyietal2016,LLNCM19}. 

(2) {\bf The `low-mass' case}, referred to as {\bf Lm}, in which the disk is ten times less massive than the nominal one. 

(3) {\bf The `low-mass, low-viscosity' case}, referred to as {\bf LmLv}, where we consider a low-mass disk in which we also consider a ten-times smaller viscosity value. In Table~\ref{table:tab1}, we give the simulation names and the main parameters.


\begin{table}
\begin{minipage}{80mm}
\caption{Disk models parameters.}
\label{table:tab1} 
\begin{tabular}{|lll|}
\hline\noalign{\smallskip}
Name  &  $\Sigma_0/ M_{\sun}r_p^{-2}$  & $\nu / r^2_p\Omega_p$   \\
\noalign{\smallskip}\hline\noalign{\smallskip}
\hline\noalign{\smallskip}
N & $6.67 \, 10^{-4}$ &  $10^{-5}$ \\
Lm &   $6.67 \, 10^{-5}$ &  $10^{-5}$ \\
LmLv & $6.67 \, 10^{-5}$  &  $10^{-6}$ \\\\
\hline
\end{tabular}

\end{minipage}
\end{table}

\begin{figure*}
\centering
   \includegraphics[width=15cm,height=17cm] {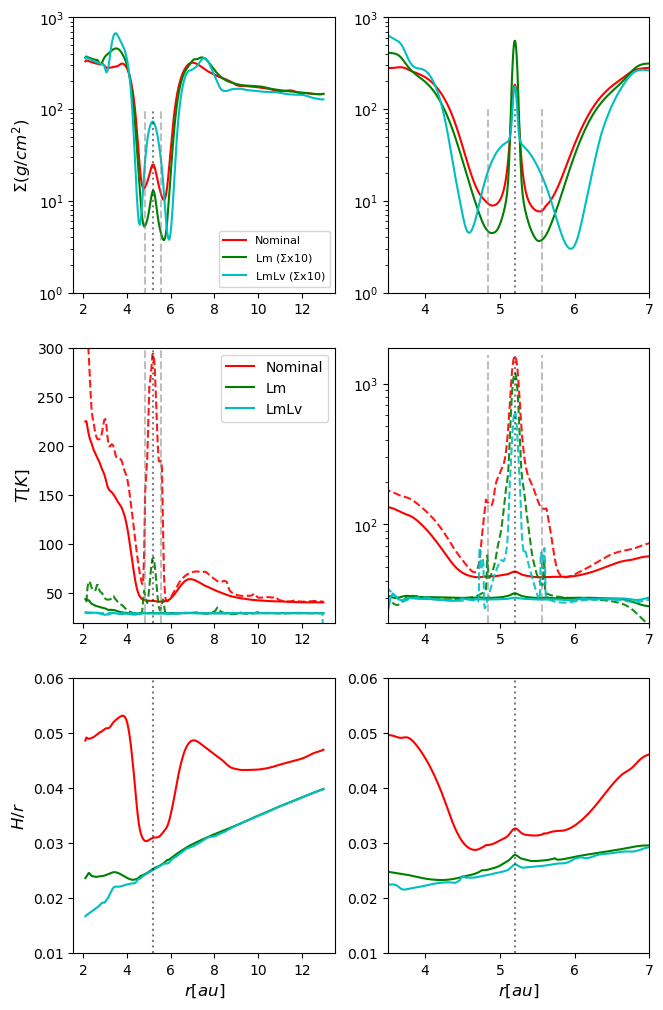}
   \caption{Radial density profiles ({\bf top}), azimuthally and vertically averaged  temperature   ({\bf middle}), and disk aspect ratio  ({\bf bottom}) profiles for the three models at the end of phase A (left) and  C (right). 
   The grey dashed lines indicate the boundary of the Hill sphere and the black dotted line the position of the planet. The density of the {\bf Lm} and {\bf LmLv} cases has been multiplied by 10 in order to superpose the density background and appreciate the differences in the gap's depth and width.
   We notice that in the {\bf LmLv} case a large amount of gas 
   is accumulated in a radial domain corresponding to the Hill sphere. This is not gas accumulated around the planet but it is the signature of the libration region around the L4 and L5 Lagrangian points not being depleted yet (a known fact in the case of low-viscosity disks).
   In the middle panel,  the midplane temperature is over-plotted with same colours (dashed) on a slice at the planet's azimuth $\varphi = 0  $.   The higher resolution associated with the deepening of the potential (decreasing smoothing) results in a huge temperature increase up to $[1600,1100,600]$~$^{\circ}K$ in the nominal model, the {\bf Lm} case, and the {\bf LmLv} model, respectively. 
In the bottom panel, we remark that it is only in the nominal case that the formation of the gap modifies  the aspect ratio considerably.}
  \label{Fig:GapComparison} 
\end{figure*}

\subsection{Disks with embedded planets: Numerical procedure}
\subsubsection{Set-up}
Before inserting the planet, we ran 2D ($r,\theta)$ simulations until the disks reach thermal equilibrium.  Thermal equilibrium is obtained with a numerical procedure described in Appendix \ref{sec:AppendixA}.
We then expanded the disk  azimuthally on $\varphi$ in the interval $[-\pi,\pi]$ and restarted our simulations from initial conditions in thermal  equilibrium.
We embedded  a planet of $20$ Earth masses in each disk model and smoothly increased (in a time interval of 20 orbits) the mass until the Jupiter mass was reached. The planet is on the midplane at $r_p = 5.2 au$ and azimuth $\varphi = 0$ (or equivalently at $(x_p,y_p)=(5.2,0)$ $[au]$) 
 on a disk of radial domain $[2,13]$ $[au]$.
\subsubsection{Planetary potential}
 The gravitational potential of the planets on
the disk ( $\Phi_p$) is modelled as in \citet{KBK09} using the full gravitational potential for disk elements having distance $d$ from the planet larger than a fraction $\epsilon$ of the Hill radius,
and a smoothed potential for disk elements with $d<r_{sm} \equiv \epsilon R_H$ according to:
\begin{equation}
    \label{KleyPotential}
\Phi _p = \left\lbrace \begin{array}{ll}
-{m_pG\over d} &  d > r_{\rm sm}, \\
-{m_pG\over d}f({d\over r_{\rm sm}}) & d\leq r_{\rm sm},   
\end{array} \right .
\end{equation}
with $f({d\over r_{\rm sm}}) = \left [ \left( {d\over r_{\rm sm}}\right)^4-2\left( {d\over r_{\rm sm}}\right)^3+2{d\over r_{\rm sm}} \right]$.
We recall that the Hill radius is defined as:
$R_H = r_p(m_p/M_{\sun})^{1/3}$. 
 We remark that (unlike the case of 2D simulations)  there is no physical need for the use of a smoothed potential in 3D simulations since the full gas column is modelled. However, the planetary potential is singular at the planet’s position and numerically handling the function $1/d$ when $d\to 0$ requires particular care. Specifically, and as shown in (\cite{LLNCM19}, Fig A.1), a minimal smoothing length of eight grid cells is required to obtain convergent results with respect to the grid resolution.
Here, we adopt the practice suggested by \cite{LLNCM19} of doubling the resolution when halving the smoothing length with a multi-phase prescription.
In the first phase   (A hereafter), we use a moderate resolution ($N_r,N_\varphi,N_\theta)=(228,680,32)$ on a uniform grid. In this way, we can benefit from the large  time step provided by the \texttt{FARGO} algorithm \cite{Masset2000}. At this stage, we have about ten grid-cells in a Hill diameter and therefore we use a large  smoothing length for the gravitational potential: $\epsilon =0.8$ $ R_H $.
After the planet's growth, we continue the simulation for additional 80 orbits to let the gap open.  Although the gaps have not converged on this integration time \citep{2019A&A...632A.118S}, the runaway growth  timescale is very short , compared to the viscous timescale required for gaps to reach their steady state.
Figure \ref{Fig:GapComparison} (left column)  shows the radial density profiles for the three models at the end of phase A. The density of the {\bf Lm} and {\bf LmLv} cases has been multiplied by 10 in order to superpose the  background unperturbed profiles and appreciate the differences in the gap's depth and width.  We remark that in the low-mass case, the gap is deeper than in the nominal one because the disk has a smaller aspect ratio (see Fig.\ref{Fig:Thr}  bottom panels). In the low viscosity  case, the gap is wider than in the other two models (as expected). \par
We continue the simulation (phase B) for additional 25 orbits using a regular grid with  resolution 
$(N_r,N_\varphi,N_\theta)=(456,1360,64)$ in the three directions. The smoothing length is smoothly reduced with respect to phase A down to $0.4$ $R_H$ . 

\subsubsection{Boundary conditions}
In the azimuthal direction, we used periodic boundary conditions;  whereas in the radial direction, we used evanescent boundaries to minimise the wave reflections \citep{deValBorroetal2006}. The reference fields (density, velocity, and gas energy)  for evanescent boundaries are those that result from the thermal equilibrium obtained running 2D $(r,z)$ axi-symmetric simulations, as described in Appendix \ref{sec:AppendixA}.
In the vertical direction at the disk surface, we considered reflecting boundaries for the velocity components. We also copied the density from the neighbours active cells and set the temperature at the values  $T_b$ (as computed in Appendix \ref{sec:AppendixA}). 
To limit the computational time,  we modelled a single disk hemisphere by using a mirror symmetry at the midplane.

\subsubsection{Increase of numerical resolution}
\begin{figure}
\centering
   \includegraphics[width=\hsize]
   {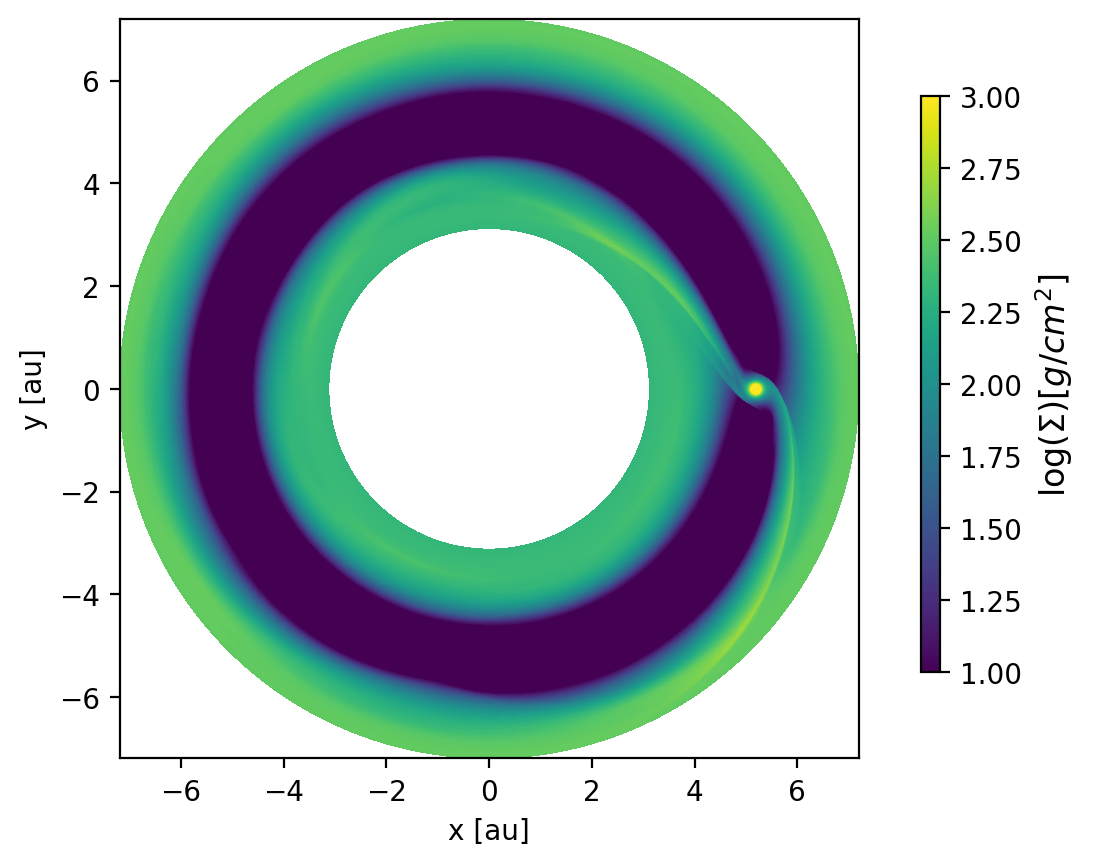}
      \caption{Vertically integrated density of the simulated nominal disk (simulation N in Table \ref{table:tab1}) in the high-resolution phase (phase C, see text) with non-uniform grid geometry. We clearly see the spiral arms originating at the planet position $(x_p,y_p)=(5.2,0)$ $[au],$  revealing the interaction between the planet and the disk. The disk annulus fully contains the gap and slightly extends over the edges.}
         \label{Fig:fulldisk} 
\end{figure}

In the next phase (C hereafter),  we used the non-uniform grid geometry (with a  number of grid-cells:  $(N_r,N_\varphi,N_\theta)=(298,1598,65)$) on a disk annulus of radial domain of
$r/r_p = [0.6,1.4]$ (or $[3.1,7.3]$ $[au]$) with $r_p=5.2$ $[au]$).
The grid has finest resolution of $\frac{\Delta r}{r_p}=0.0013$ in  the three dimensions, close to the planet's position.
The spacing is quasi-constant in the Hill sphere, increases quadratically  and  has a maximum resolution ratio of 6 in the radial direction (with 10 and 2.5 in the azimuthal and in  the vertical directions, respectively, not shown). The number of grid-cells in a Hill radius is of 50. We  discuss the convergence of results  with respect to resolution and smoothing length  in \ref{sec:AppendixB}. The integration annulus fully contains the gap and the radial domain for evanescent boundary conditions,
as seen in Fig.\ref{Fig:fulldisk}.


The reference fields for the evanescent boundaries are the gas fields inherited by those computed  at the
end of phase B. Because of the non-uniform geometry of the grid in azimuth, we cannot use the \texttt{FARGO} algorithm in phase C; thus, the simulations take a longer time and we ran them for an additional ten orbits.  The smoothing length in phase C is  reduced with respect to phase B down to $\epsilon = 0.2$ $R_H$ in 1 orbital period { (corresponding to 10 grid-cells in the smoothing lenght)}. All of the figures in the following are snapshots at five orbital periods out of a total of ten orbital periods for this phase. 
From phase A to phase C, we increased the resolution and reduced the smoothing length: more gas is accreted  in the planet vicinity (Fig.\ref{Fig:GapComparison}, top right panels). 
In  Fig.\ref{Fig:GapComparison} (middle-right panel),  we observe averaged temperatures similar to the results of phase A on the coarsely grained grid, along with a huge increase (of a factor of 5-20) in the temperature at the planet's position. This increase in temperature near the planet follows from the higher spacial resolution, along with the deeper potential and the fact that the main source of heating is compression \citep{LLNCM19}. The midplane temperature suddenly drops at the boundary of the Hill sphere; however, small temperature spikes are observed near this boundary in the three models due to the intersection of the selected (azimuthal) slice  with the spiral wake (as we show in the next sections).
 We can already see from this study that local properties of temperature and density in the Hill sphere of a giant planet that has opened a gap depend on such disks parameters as the mass and the viscosity.


\begin{figure*}
\centering
\includegraphics[width=\textwidth,height=22cm]{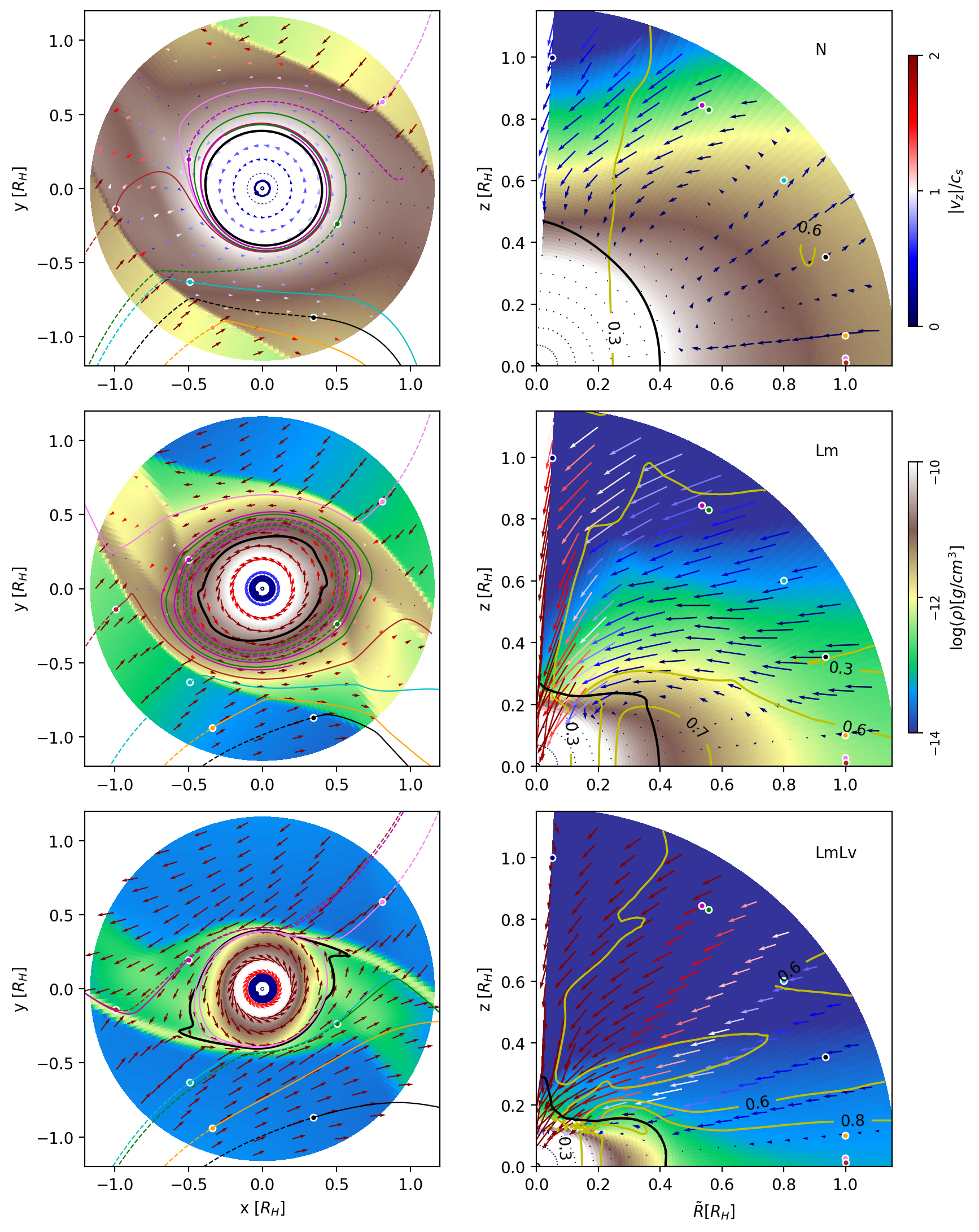} 
   \caption {Parameter details for all three models.  
{\bf Left}: Midplane density field interpolated in cartesian planeto-centric coordinates$(x,y)=(R\cos(\phi),R\sin(\phi))$ in units of the Hill radius). The cartesian components $(v_x,v_y)$ of the midplane velocity field are plotted with arrows. The coloured lines are straightforward (full) and backward (dashed) integration of streamlines obtained with the 2D midplane velocity field.
   {\bf Right}: Azimuthally averaged density field interpolated in planeto-centric coordinates
   and represented in $(\tilde R,z)=(R\sin(\vartheta),R\cos(\vartheta))$ plane. 
   The azimuthally averaged and density weighted velocity is represented  with arrows.
   The colours of the arrows  correspond to values of the of the gas speed normalised over the sound speed(respectively the midplane speed (left column) and the absolute value of $v_z$  (right column)).
   The yellow contour levels correspond to the vertical component of the angular momentum in units of the Keplerian momentum with respect to the planet. 
   The  coloured markers  give the initial conditions for the streamlines of Table \ref{table:tab2},
   projected  on the $(x,y)$ and
   $(\tilde R,z)$ planes, respectively,  and the black solid lines identify the 0 energy contour (see Eq. \ref{energy})} 
\label{Fig:SliceSpherical} 
\end{figure*}

\begin{figure*}
\centering
\includegraphics[width=\textwidth]{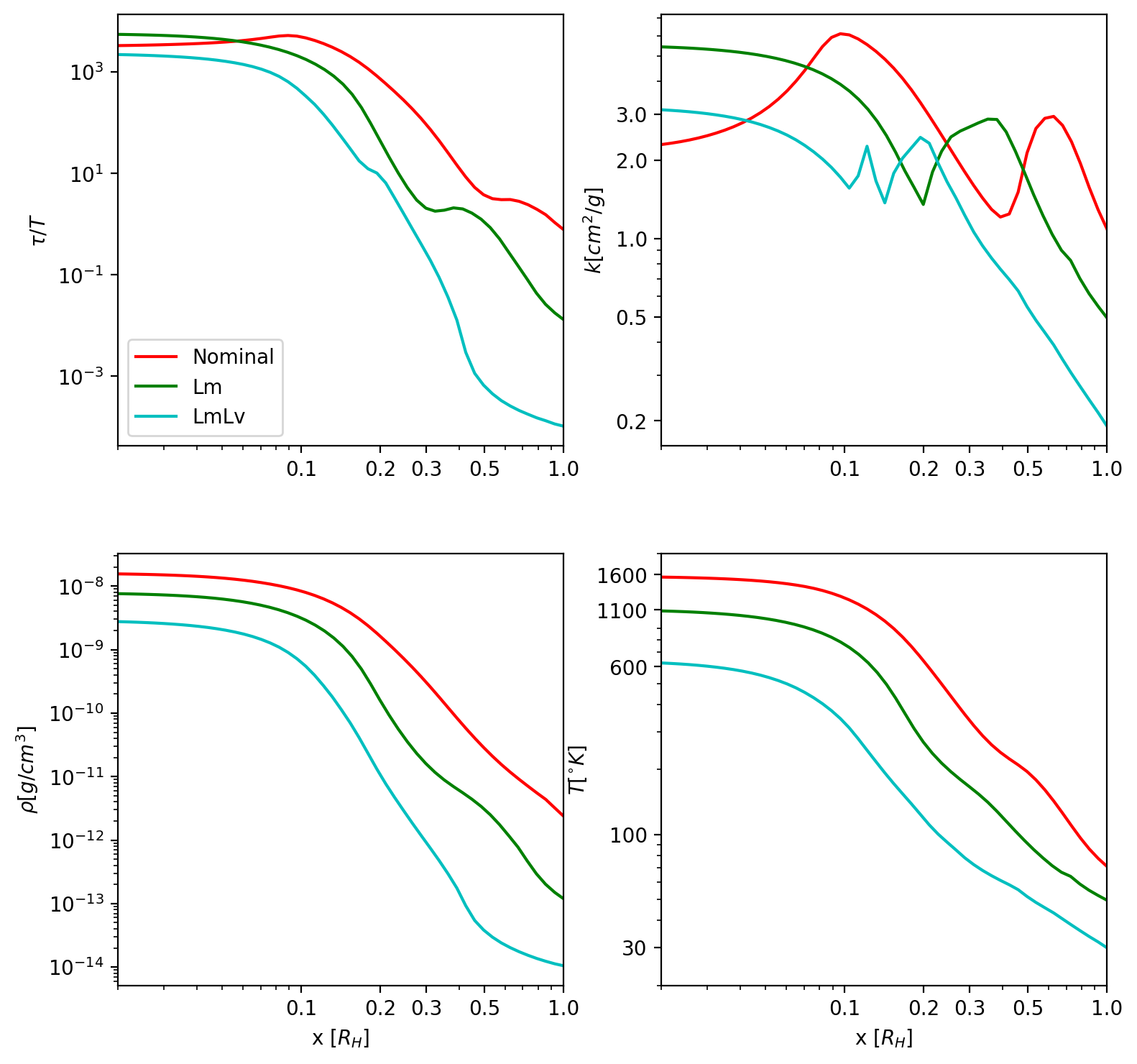} 
   \caption {Parameter details for all three models. {\bf Top-left:} Radiative diffusion timescale
   (from Eq. \ref{Eq:tau}).   {\bf Top-right:} opacity. {\bf Bottom-left:} Volume density. {\bf Bottom right:} Temperature.
   All the quantities are obtained through averages over spherical shells centred on the planet.}
\label{Fig:Shells} 
\end{figure*}

\section{Gas dynamics in the Hill's sphere }
\label{Sec:Dynamics}
In this section, we provide a detailed description of the dynamics in the planet's vicinity
and quantify the amount of gas that flows through the Hill sphere as well as the accretion rate in the Hill sphere. In a follow-up paper (\cite{Cridland2024}, submitted), we will present the chemical evolution of gas along streamlines computed in this work. 

To focus on the dynamics in the planet's vicinity, we  worked in the planetocentric reference frame and used either Cartesian coordinates $(x,y,z)$ 
or spherical coordinates ($R,\phi,\vartheta),$
with $R$ representing the radial distance of the planet, $\phi$ the azimuthal angle in $[0:2\pi]$, where the x-axis, namely, $\phi=0$, marks the Sun-planet direction;
$\vartheta$ is the colatitude in the interval $[0:\pi/2],$
with $\pi/2$ indicating the planet's equatorial plane (corresponding to   the disk's midplane) and $0$ indicating the pole of the sphere. 
In Fig.~\ref{Fig:SliceSpherical},  we provide the density field at the disk midplane (left panels)
and the azimuthally averaged temperature field (right panels) represented on the plane 
$(\tilde R,z) \equiv (R \sin(\vartheta),R\cos(\vartheta))$ 
for the three models {\bf N} (top), {\bf Lm} (middle) and {\bf LmLv} (bottom).
The velocity field $(v_x,v_y)$ is over-plotted on the background midplane density with white arrows. 
On the right panels, the velocity field is azimuthally averaged and density weighted as follows:
\begin{equation}
\label{Eq:densweight}
    <X>_{\rho} ={\frac{\int_0^{2\pi} \rho X d\phi} {\int_0^{2\pi} \rho d\phi}}
,\end{equation}
where $<X>$ accounts for both components
$(<v_{\tilde R}>,<v_z>)$. 
The colours of the arrows correspond to the  vertical component $<v_z>_{\rho}$ 
in units of the sound speed.

\begin{figure}
\centering
\includegraphics[width=\hsize]{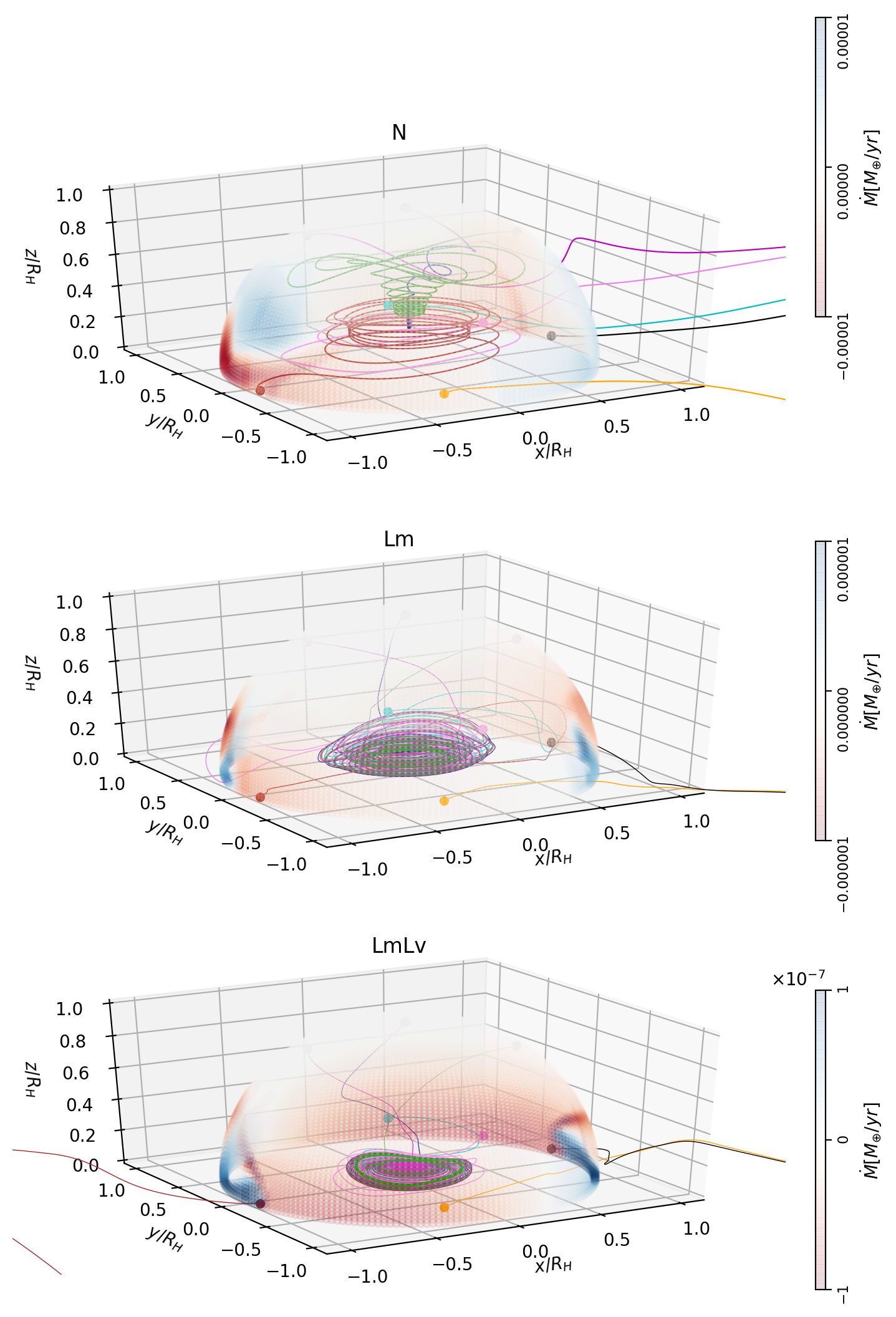}
 \caption {3D representation of streamlines with initial conditions on the Hill sphere (see Table \ref{table:tab2}). The colour bar corresponds to planeto-centric mass flow rate on the Hill surface. The sign is negative for mass inflow and positive for mass outflow. The total integration time is short (about four Jupiter orbits) in order to make this plot clear. The evolution of the same streamlines on a longer integration allows us to follow the dynamics on a projected plane $(x/R_H,y/R_H)$  and to appreciate the evolution of the distance with respect to the planet in Fig.\ref{Fig:Projection}}
\label{Fig:Streamlines} 
\end{figure}

\subsection{Midplane views}
 On the plot displaying the density midplane, we clearly see the spiral wake as an over-dense region that shrinks and becomes very thin when moving from the nominal to the {\bf LmLv} case. The central massive region (envelope) extends to about 0.5 $R_H$, gas is subsonic inside 
 the envelope; whereas it approaches the wake at supersonic speed and is then deviated and slowed at subsonic speed after shock with the wake. 
 We remark that the density structure is similar to the one described in (\cite{Schuliketal2020}, see their Fig. 3) for similar protoplanetary disk initial conditions (same viscosity and twice the smaller surface density at the planet's location), but with a small opacity value of $0.01 g/cm^2$.  The wake structure seems not to depend on the opacity, however, the region inside 0.4 $R_H$ is in our case almost not rotating. This is typical of pressure-supported envelopes and by  lowering the opacity, \cite{Schuliketal2020}  found a rotationally supported structure inside 0.4 $R_H$ (and  strong spiral wakes in the CPD).  
 In the middle and bottom panels ({\bf Lm} and {\bf LmLv} cases), in addition to the spiral wake, we observe additional spiral arms: this is a sign that in these cases the gas surrounding the planet forms a rotationally supported structure or CPD with its tidal spiral arms.
When the gas approaches the proto-planet, its motion is deviated by the gravity of the planet. From the literature, we have the definition of three regions \citep{2008ApJ...685.1220M,2010MNRAS.405.1227M,2012ApJ...747...47T} corresponding to classical circulation
(also called a pass-by region) and horseshoe motion at distances of $~>R_H$, along with a third region for distances of $R<R_H$ (called atmospheric region),  where the gas is highly perturbed by the planet. 

{ Inside the Hill sphere, we can identify the region where the gas is bound to the planet through energy considerations. Actually, the gas escapes the planet's  potential  well if its total energy is larger than the effective potential at the Hill radius: $\Phi _H = -Gm_p/R_H -3/2\Omega_p^2R_H^2  $ ;  here, the second term is the centrifugal contribution in the local approximation (Hill's equations).
Therefore, we can say that the gas is bounded where $E\leq 0$ with the specific energy defined as:
\begin{equation}
\label{energy}
E = \frac {v^2}{2}+c_vT+\Phi_p-\frac{3}{2}\Omega_p^2x^2-\Phi_H
.\end{equation}
The zero-level energy contour (black line in the left panels of Fig.~\ref{Fig:SliceSpherical} ) shows that the  gas is bounded to the planet  up to a distance of about 0.4 $R_H$, with a spherical geometry in the  {\bf N} case and with an elongated or disk like structure on the {\bf Lm} and {\bf LmLv} cases.} 
\par
In the left panels of  Fig.~\ref{Fig:SliceSpherical}, we have also plotted forward (full lines) and backward (dashed lines) integration for the set of initial conditions of Table \ref{table:tab2} projected on the midplane ($z=0$), using the 2D velocity fields components  $(v_R,v_{\phi})$. Ignoring the vertical motion, the flow pattern 
emphasises the influence of the wake on the gas motion.
In the top panel (nominal case), after being deviated by the shock with the spiral wake, the brown and pink lines according to the velocity field are further deflected near the Lagrangian point, $L_1$, at approximately $x=-0.6,y=0.1$  $R_H$, and
reach the boundary of the hot envelope at about 0.4 $R_H$. On average the radial motion of the midplane is directed inward, gas moves towards the hot central envelope and, in the full 3D case (Fig. \ref{Fig:SliceSpherical}, top-right) is then lifted up. A consequent depression in the midplane will keep the gas moving toward the hot envelope ("green" and "magenta" streamlines), while 
the others ('cyan', 'black', and 'orange') pass further away from the planet and 
have a typical horseshoe motion.

In the {\bf Lm} case, the brown and pink lines are deviated by the wake, have about half a rotation around the planet and are bent again  by respectively the CPD spiral arm and the wake and finally leave the Hill region, while the green and magenta lines spiral out and are kinked at each rotation by both the wake and the CPD spirals. A similar behaviour is observed in the {\bf LmLv } case,  with the majority of streamlines entering the  Hill sphere participating to horseshoe motion.

\par
\subsection{Vertical views}
The contour levels plotted in Fig. \ref{Fig:SliceSpherical}, right panels 
correspond to the vertical component of the specific angular momentum, $j$, in a non rotating frame centered on the planet,  normalised  over the Keplerian angular  momentum relative to the planet. More precisely, we have:
\begin{equation}
    \label{angmom}
    j =\frac{ <v_{\phi}\tilde R>_{\rho} +\Omega_p\tilde R^2}{\sqrt {Gm_p \tilde R}}
,\end{equation}
with $\tilde R$ being the cylindrical radius.
In the nominal model, along the polar directions, the gas has less than $30\%$ of the Keplerian angular momentum  and flows quasi-vertically towards the planet with a subsonic sound speed -- as previously reported in the adiabatic approximation \citep{2019ApJ...887..152F,2019MNRAS.488.2365B}.  The in-fall speed further decreases when the gas approaches the hot envelope at about $z \sim 0.5  [R_H]$. The gas { is bound} to the planet inside a spherical envelope   of radius $0.4$ Hill radii ({   black line showing the zero energy contour, according to Eq. \ref{energy} in} Fig. \ref{Fig:SliceSpherical}, top-right panel).
This  is typical of a pressure-supported envelope  in agreement with the finding of \cite{Szulagyietal2016,LLNCM19,2019ApJ...887..152F}. 
Within the  envelope, the gas speed is negligible with respect to the values observed at larger  distances.

When looking at the {\bf Lm}, {\bf LmLv} cases, we see that   gas flows quasi-vertically at supersonic speed
in the region bound by the line with 0.3 Keplerian angular momentum
in the {\bf Lm} case and on a larger fraction of the Hill sphere in the
{\bf LmLv} case.
In the {\bf Lm} case,  we observe a dense quasi-spherical envelope of a 
radius of $\sim 0.2 [R_H] $  
and a thick disk  that is partially supported by pressure. The separation between the envelope and
the surrounding disk can approximately  be set at $\sim 0.2 [R_H] $: for larger distances,  the angular momentum gets larger than  $60\%$ of the Keplerian value. In the {\bf LmLv} case, the rotational support increases to  $80\%$ of the Keplerian one at a distance of
$\tilde R>0.2 [R_H]$ . 
In the{  region bounded to the planet and delimited by the zero energy contour (black line in the right panels of Fig. \ref{Fig:SliceSpherical}} ),  the averaged motion has negligible velocity with respect to the rest of the Hill sphere.
The disk structure appears to shield the gas entering in the Hill sphere at low altitude by deviating it, lifting it up ( middle right panel of Fig.~\ref{Fig:SliceSpherical}) above the disk structure,  and then moving it\ radially towards the planet, where it falls almost vertically when the angular momentum is low.
We describe the dynamics in the planetary system in  Sect. \ref{Sec:Streamlines}.

\subsection{Rotational support}
The fraction of pressure support with respect to rotational  support of a planetary system  depends on the cooling efficiency in the Hill sphere \citep{2019ApJ...887..152F}.  \cite{2009MNRAS.397..657A} had previously shown that the rotational support increases  when decreasing the opacity and this result was recently confirmed in a detailed study by \cite{Schuliketal2020}.
Here, we use the Bell \& Lin opacity $k(\rho,T)$ for our three models with a dust-to-gas ratio of $0.01$.
{ Although the inner and denser part of the planetary envelope may be convective \citep{2021MNRAS.508..453Z}, in the outer part, or in the CPD forming region, cooling is mainly radiative \citep{2019ApJ...887..152F,2024arXiv240214638K}.}
Radiative cooling can be quantified by computing the timescale over which radiation is diffused over the Hill sphere. Following the definition of  \cite{2010A&A...523A..30B} we compute:
\begin{equation}
    \label{Eq:tau}
    \tau = \frac {R_H^2} {D_r}
,\end{equation}
where $D_r$ is a radiative diffusion coefficient in the optically thick limit:
\begin{equation*}
    D_r = \frac {4caT^3} {3c_v\rho^2k(\rho,T)}.
\end{equation*}

In Fig. \ref{Fig:Shells} we show, as a function of the radial distance from the planet, quantities averaged on spherical planetocentric shells, according to:

\begin{equation}
    \label{Eq:shellaverage}
    F(R) = \frac{1}{2\pi R^2}\int _{0}^{2\pi} \int _{0}^{\pi/2} f(R,\theta,\phi/2)R^2 sin(\vartheta) d\vartheta d\phi
,\end{equation}
where the function, $f$, is the cooling timescale in units of orbital periods $\tau/T$ (top-left), the opacity (top right), and the volume density and temperature (bottom panels). \par
The inner region up to 0.1 $R_H$ has a high density and  opacity values larger than 1 $cm^2/g$ for the three models (see Fig. \ref{Fig:Shells}) with a moderate decrease of the opacity inside 0.1 $R_H$, where the temperature is higher than $1000 ^\circ$ and the opacity decreases because of gas ionisation processes. The radiative cooling timescale is of the order of $10^3$ orbital periods for the three models.  At distances greater than 0.3 $R_H$ from the planet, $\tau$ decreases to values smaller that the dynamical timescale for both the {\bf Lm} and {\bf LmLv} cases. In a recent paper, \cite{2024arXiv240214638K}
 proposed a necessary condition for the formation of  a CPD depending on the value of the { radiative} cooling timescale. 
According to this condition, a CPD forms when the cooling time is of  the  same order or smaller   than the orbital period at the tidal truncation radius: $\simeq 0.3 R_{H}$. Although we do not have  the same definition of $\tau$ as in  \citep{2024arXiv240214638K}, this condition is fullfilled by the {\bf LmLv} case and marginally fulfilled in the   {\bf Lm}  case where the disk structure is thick and partially supported by pressure.

\subsection{Streamlines}
\label{Sec:Streamlines}
A detailed description of gas dynamics in the planet's vicinity is given via the description of streamlines in \cite{2008ApJ...685.1220M,2010A&A...523A..30B,2012ApJ...747...47T} in the case of isothermal and adiabatic models. 
 With the aim of comparing the gas dynamics in our three models.
 we consider in this section a set of initial conditions on the Hill sphere (see Table \ref{table:tab2} and Fig. \ref{Fig:SliceSpherical} for projections on the disk midplane and on the vertical slice).
\par
Streamlines give the motion of gas parcels for
quasi-stationary velocity fields.
We checked this requirement by comparing the velocity fields at different snapshots on
our hydrodynamical integration,  which cover nine orbits of Jupiter around the star. For the streamlines, we integrated the velocity field  up to ten Jupiter orbits forwards and to two Jupiter orbits backwards. 
Figure~\ref{Fig:Streamlines} reports the  streamlines on a portion of the space slightly larger than the Hill sphere for an integration time of about four Jupiter orbits.  On the Hill sphere, we have plotted the radial gas flow rate in planetocentric coordinates: negative values correspond to gas inflow  and positive values correspond to outflow  (red and blue points, respectively, in Fig.~\ref{Fig:Streamlines}). 
From a simple consideration of the flow sign, we see that gas inflow dominates quasi totally in the {\bf Lm} and {\bf LmLv} cases (red regions), whereas a large fraction of the sphere is dominated by outflow (blue regions) in the nominal case. We provide quantitative values of inflow and outflow rate on the Hill sphere as a function of $z/R_H$ in Sect.
\ref{Sec:Accretion}.

\clearpage
\newpage
\begin{figure}
\centering
\includegraphics[width=\hsize]{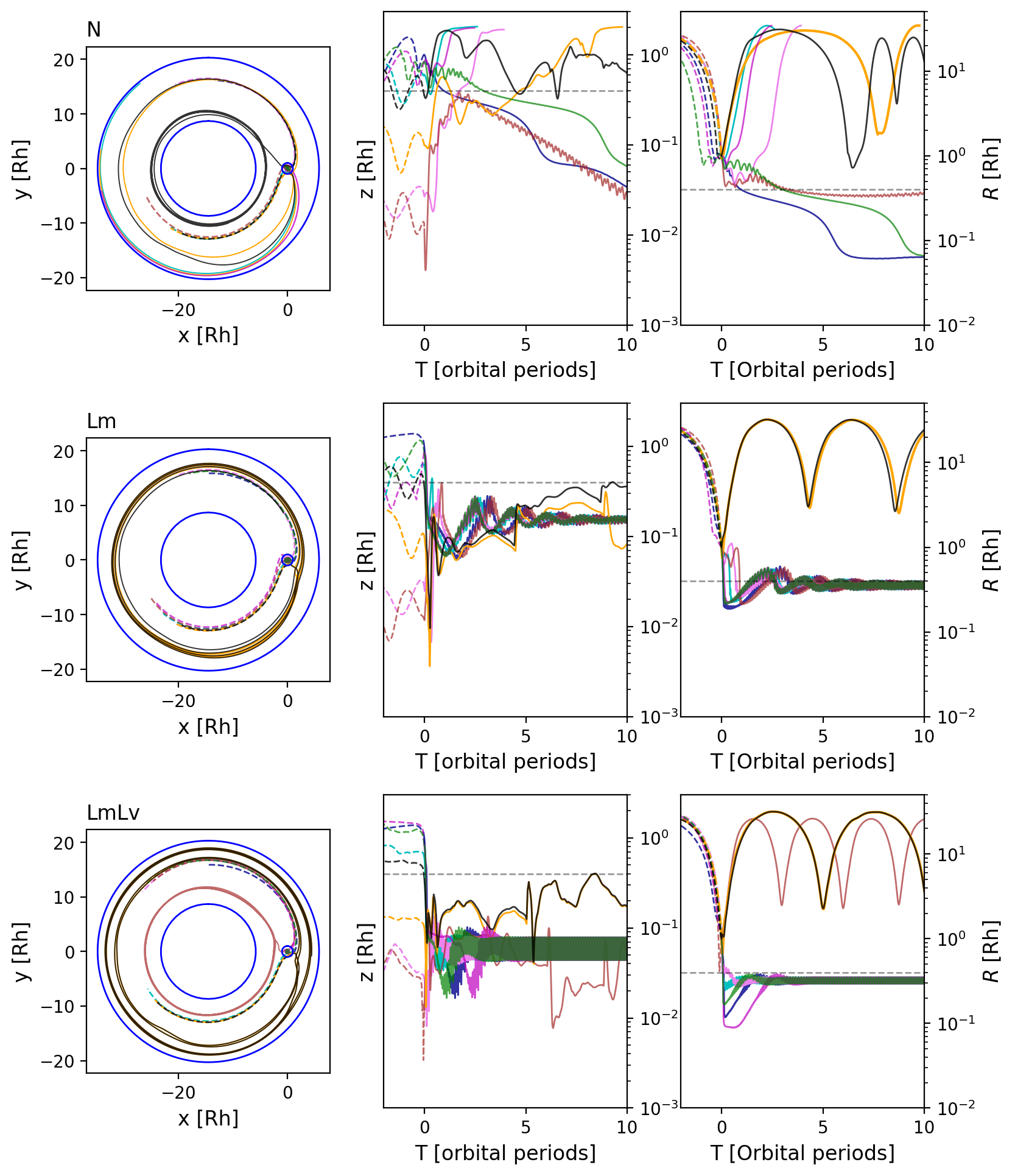}
 \caption {Projection of the streamlines of Fig. \ref{Fig:Streamlines}  on the planetocentric $x-y$ plane (left). The blue contours give  the boundaries of the disk. Gas approaching  the Hill sphere can have a rapid transit phase in the horseshoe region and be released in the inner or the outer disk trapped in the planetary system. Vertical evolution of  streamline with respect to time (middle). Evolution of the distance with respect to the planet as a function of time (right). The dashed lines correspond to backward integration: in the left panel we see that gas approaching the Hill sphere comes from U-turn motion passing close to the planet  from the inner or outer branch of the horseshoe region. }
\label{Fig:Projection} 
\end{figure}

\begin{table}
\begin{minipage}{90mm}
\caption{Initial conditions for the streamlines plotted in Figs.~\ref{Fig:Streamlines} and \ref{Fig:Projection}. }
\label{table:tab2} 
\begin{tabular}{|lll|}
\hline\noalign{\smallskip}
Color  &  Initial conditions ($x/R_H,y/R_H,z/R_H)$  & \\
\noalign{\smallskip}\hline\noalign{\smallskip}
\hline\noalign{\smallskip}
dark blue & $(-0.041337,-0.028092,+0.998750)$ & \\ 
magenta & $(-0.497865,+0.197119,+0.844556)$ & \\ 
green & $(+0.504023,-0.237175,+0.830487)$ &\\ 
cyan &   $(-0.489703,-0.631321,+0.601352)$ & \\ 
black & $(+0.344279,-0.869548,+0.354059)$ & \\  
orange & $(-0.338603,-0.940575,+0.100000)$ & \\
pink & $(+0.808720,+0.587569,+0.027115)$ & \\ 
brown & $(-0.9902741,-0.139131,+0.01199)$ & \\ \\ 
\hline
\end{tabular}

\end{minipage}
\end{table}

As a first global comment, we can observe that in the {\bf N} case gas entering at high latitude tends to remain trapped and  rotate around the planet, either exiting at high altitude or slowly approaching the planet's envelope. In the other two cases, a larger number of streamlines are trapped in the planetary system and a more flattened disk structure is observed for the {\bf LmLv} case, with respect to {\bf Lm} one. We remark that a larger internal energy budget inside the Hill sphere as for the {\bf N} case (see Fig.\ref{Fig:Shells} bottom panels) seems to have a key role in determining the fraction of  gas accreted in the planetary region with respect to the total gas entering in the Hill sphere. We {  provide a quantitative study} in Sect. \ref{Sec:Accretion}.

\par 
In the following, we describe the behaviour of single streamlines by also looking  at quantities reported on Fig.~\ref{Fig:Projection}, where we can appreciate the projected path of streamlines on the $x-y$ global disk as well as  the evolution with time of  their $z$ and radial
components relative to the planet.  The dashed lines on the left panels of   Fig.~\ref{Fig:Projection} correspond to backward integration:  we see that in all cases, the gas approaching the Hill sphere comes from U-turn motion passing close to the planet; namely, close to Lagrangian points L1 and L2,  from the inner or outer branch of the horseshoe region (in agreement with the findings of
\cite{2008ApJ...685.1220M,2010MNRAS.405.1227M}.
\par
\subsubsection{Nominal case}
In the nominal case, the  initial conditions  that are very close to the planet's pole ('dark blue' and 'green' streamlines)  gas enter deep into the envelope and {is }accreted (see also Fig.~\ref{Fig:Projection}, top-middle panel), although very low density is carried by gas at high $z$ values.  We consider that the gas carried by a streamline  is accreted when it enters the region gravitationally bound to the planet (shown by the black lines in Fig. \ref{Fig:SliceSpherical}).
Not all of the gas entering the Hill sphere close to the planet's pole is accreted,  in fact we can see that the 'magenta' line, which has an initial value of $z/r_H$ very close to the 'green' one, but  different $x$ and $y$ positions (see Fig.~\ref{Fig:SliceSpherical}, left panel) escapes from the Hill sphere. Precisely, in this case, the gas has a short residence in the Hill sphere moving at lower z values and then is lifted higher up and leaves the Hill sphere reaching the outer disk region (Fig.~\ref{Fig:Projection}, top row). The computation stops when a streamline reaches the computational domain's boundaries.
Starting further down at $z/r_H =0.6$ and $z/r_H =0.35 $  (respectively, 'cyan' and 'black' streamlines), the gas is lifted up in $z$, according to the direction of the vectors in Fig. \ref{Fig:SliceSpherical} ( top-right panel). In the longer integration presented in Fig. \ref{Fig:Projection}, top panels,
 the gas transits to the  outer disk ('cyan' streamline),
 whereas it re-enters the Hill sphere ('black' streamline) after a horseshoe orbit and, finally, transits  into the inner disk
(Fig.~\ref{Fig:Projection}, top left panel).
 A similar behaviour occurs with respect to the 'orange' 
 streamline starting at $z/r_H =0.1$: the gas makes a 
 U-turn in the Hill region (similar to the midplane streamline of Fig.~\ref{Fig:SliceSpherical}, top-left) followed by an horseshoe orbit, then again approaches   the Hill sphere, entering at higher z values at about 
 $t=8$ orbital periods; finally, it leaves the horseshoe region and reaches the disk's upper boundary (see Fig.~\ref{Fig:Projection}, top panels).

\begin{figure}
    \includegraphics[width=\hsize]{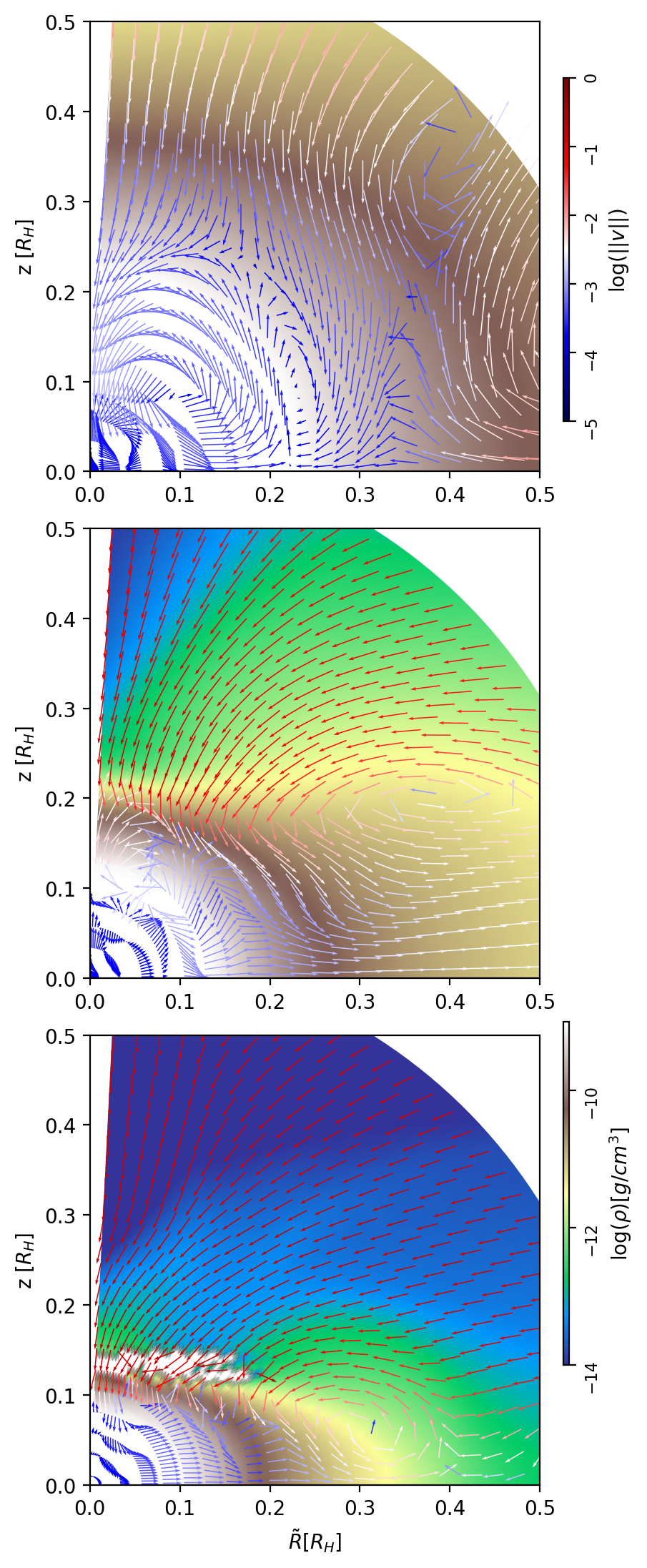}    \caption{Same as Fig. \ref{Fig:SliceSpherical}, right panels showing a zoom close to the planetary system. Vectors of components $(v_{\tilde R},v_z)$ are normalised and the colours correspond to the norm values.
The norms cover a range of about five orders of magnitude, with smaller values deep inside the planet's envelope.}
    \label{Fig:VectZoom}
\end{figure}
 
 For streamlines with initial conditions that are very close to the midplane, we observe:  a close approach to the bounded region at $d \simeq 0.4 R_H$ and a few rotations around the planet that are similar to the corresponding midplane
 streamline
 ('pink' line). Then the gas is lifted up  and transits to  the outer disk remaining at large z values (Fig.~\ref{Fig:Projection}, top panels). Gas starting on the midplane, but at a $x$ $y$ position 
 close to the L1 lagrangian point ('brown' streamline), is deflected and remains bounded at a distance of about 0.4 Hill radii; namely, at the  planet's envelope boundary (Fig. \ref{Fig:Projection}, top-right panel).
We notice that  the dynamical behaviour is similar to the one observed for the
 corresponding midplane streamline of  Fig. \ref{Fig:SliceSpherical} (top left); however, in the full 3D velocity field, the gas is lifted up to about 0.4 Hill radii  (Fig.~\ref{Fig:Projection}, top middle panel) and then oscillates vertically whilst rotating around the planet (as indicated by the small period oscillations in Fig.~\ref{Fig:Projection}, top ( middle and right panels).  If the gas entering in this case from the midplane carries small dust particles, it will then pollute the planet's envelope. 
 \subsubsection{The {\bf Lm} and {\bf LmLv} cases}
For the other two cases, the dynamics appear very different:  the majority of streamlines are trapped in a toroidal disk structure  with an outer bound at about 0.4 $R_H$ (slightly larger than the tidal truncation radius) and an inner bound at about  0.25 $R_H$. To explain  the path of such streamlines, we show the velocity field for distances $<0.5 R_H$ in 
the $(\tilde R,z)$ plane
with normalised vectors (Fig.\ref{Fig:VectZoom}). The vector colours correspond to the norm of $(v_{\tilde R},v_z)$.
The norm of the vectors cover a range of about five orders of magnitude, with smaller values inside the planet's envelope where the gas is bounded.
Gas that shocks the envelope  ('dark blue', 'green', and 'magenta' streamlines) at about $z\sim 0.2$ and $z\sim 0.12$ (for the ({\bf Lm} and {\bf LmLv} cases, respectively) is deviated and trapped in circulating motion in the  $(\tilde R,z)$ plane (toroidal motion in the full 3D field)
 .  Initial conditions making a U-turn far from the planet as the 'yellow' and 'black' streamlines undergo rapid transit, as observed for the corresponding midplane streamlines of Fig.~\ref{Fig:SliceSpherical} (middle- and bottom-left panels). 
The 'brown' streamline  is deflected by the wake and enters in the
inner protoplanetary disk  in the {\bf LmLv} case, whereas it is captured in the disk structure in {\bf Lm} case (Figs.~\ref{Fig:Streamlines} and~\ref{Fig:Projection}, middle and bottom panels). When we observe the 'brown' streamline, as computed with the midplane velocity field in the {\bf Lm} case, we see two important deviations when the gas encounters  the dense spiral wake, which eventually makes it  leave the Hill sphere. 
Using the full 3D velocity field, the encounter with the wake also modifies the motion  in the vertical direction and lifts up the gas (as described in \cite{2019A&A...632A.118S}, Fig.14, and explained as baroclinic kicks). 
At the encounter with the outer wake,  the streamlines are deviated again and  appear to be finally trapped in the disk region.   A similar behaviour is observed for the 'pink' streamline in the {\bf Lm} case.
Despite the complexity of 3D motion in the Hill sphere, we can conclude by saying that the accretion of gas entering in the Hill sphere is favoured in the {\bf Lm} and {\bf LmLv} cases. Here, the gas is colder outside the planet's envelope  with respect to the {\bf nominal} case, which has a larger internal energy budget inside the Hill sphere (see also Fig.\ref{Fig:Shells}, bottom panels).

\begin{figure}
\centering
\includegraphics[width=\hsize]{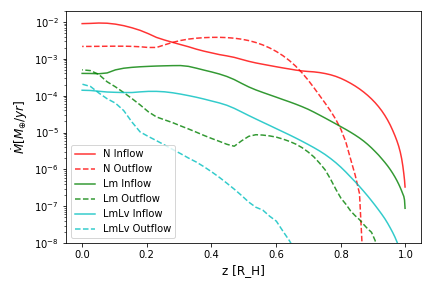}
 \caption {Absolute value of the inflow and outflow mass rate as a function of z on  the Hill sphere. In the {\bf Lm} and {\bf LmLv} cases the mass inflow is larger than the mass outflow except very close to the midplane which appears decreating. Instead, in the {\bf N} case, the Hill sphere is accreting up to $z=0.2 [R_H]$ and for $z>0.7 [R_H]$. In all the models,  the contribution to the total mass flow in the Hill sphere becomes negligible at $z>0.8 [R_H]$
 }
\label{Fig:ZAccr} 
\end{figure}

\begin{figure}
\centering
\includegraphics[width=\hsize]{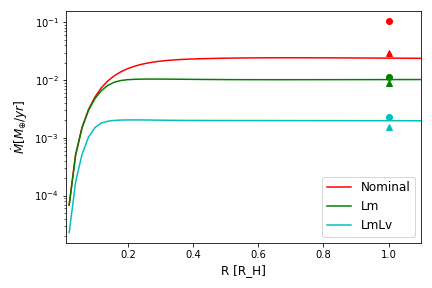}
 \caption {Accretion rate as a function of the distance to the planet in the Hill sphere computed interpolating the time series of the  mass contained in a sphere of radius $R$ over 9 planet's orbital periods (see text). For each case, we report the rate of inflow (circles) and the accretion rate on the Hill sphere (triangle), as computed on a snapshot at five orbital periods (see Table \ref{table:tab3}). }
\label{Fig:Accr} 
\end{figure}

\subsection{Accretion}
\label{Sec:Accretion}
From the values of mass flow rate (plotted on the Hill sphere in Fig. \ref{Fig:Streamlines}), we have a qualitative 
view of regions of gas inflow and outflow. We sum  these values on the azimuthal direction  to provide 
the  total inflow $\dot M_i(z)$ and outflow mass flow rate $\dot M_o(z)$  as a function of $z [R_H]$. \par 
The resulting values are plotted in Fig. \ref{Fig:ZAccr}, taking the absolute value for the inflow rate. The difference between the inflow and the outflow mass rate will give the net accretion rate, simply called the 'accretion rate' in the following. In the {\bf Lm} and {\bf LmLv} cases (Fig. \ref{Fig:ZAccr}), the mass inflow is  larger than the mass outflow except very close to the midplane. In other words, very close to the midplane the flow is decreating and the rest of the Hill sphere is accreting. Instead, in the {\bf N} case, the Hill sphere is accreting up to $z=0.2 [R_H]$ and for $z>0.7 [R_H]$. In all the models,  the contribution to the total mass flow in the Hill sphere becomes negligible for $z>0.8 [R_H]$. 
The total mass flow in and out of the Hill sphere and the resulting accretion rate  are reported in  $[M_{\oplus}/yr]$ in Table\ref{table:tab3}.

\begin{table}
\begin{minipage}{90mm}
\caption{Total mass inflow, $|\dot M_i|,$ and outflow, $\dot M_o$, rates.}
\label{table:tab3} 
\begin{tabular}{|llll|}
\hline\noalign{\smallskip}
 Model & $|\dot M_i|$  & $\dot M_o$  & $\dot M [M_{\oplus}/yr]$ \\
\noalign{\smallskip}\hline\noalign{\smallskip}
\hline\noalign{\smallskip}
N  &0.104 & 0.075 & 0.029 \\
Lm & 0.0115 &  0.0022  &  0.009  \\
LmLv &0.0023 & 0.00074 & 0.00156  \\ 
\hline
\end{tabular}
\tablefoot{
  The data are obtained from the sum of the values $\dot M_i(z)$ and  $\dot M_o(z)$ plotted in Fig. \ref{Fig:ZAccr}. The resulting values have been multiplied by 2 to take into account the whole Hill sphere.
The fourth column is simply the difference between the inflow and the outflow rates and  provides the net  accretion rate on the Hill sphere.
The computation is done on a snapshot at five of the planet's orbital periods. 
 }

\end{minipage}
\end{table}

We can now quantify the accretion rate inside the Hill sphere
using the whole time series of data at our disposal.
More precisely, we first compute the total mass $m_{R}(t)$ contained on   semi-spheres  of radius, $R,$ centred on the planet with $R$ going from the 0.01 to 1 Hill radius.  We have a total integration time of  ten orbits  (phase C; see Sect. \ref{Sec:model}) and  the first orbit is used to smoothly reduce the smoothing length.  Therefore, we obtained the mass accretion rate from a linear  interpolation of the function  $2m_{R}(t)$ (where the factor of 2 gives the whole mass on the sphere)  over the remaining nine orbits. In Fig.~\ref{Fig:Accr},
we plot the accretion rate in Earth masses per year.
At the Hill radius, we also plot the total rate of inflow and the accretion rate from Table \ref{table:tab3}.
We observe the good agreement among the accretion rates computed through radial flow on a single time snapshot and the one obtained through interpolation over nine orbits.
\par
In Fig.~\ref{Fig:Accr}, we notice that  below 0.3 Hill radii for the nominal case (0.2 and 0.1 Hill radii for  the {\bf Lm} and {\bf LmLv} cases, respectively),  the accretion rate suddenly decreases and becomes negligible when approaching the planet.
We can consider the interior envelope to be slowly contracting and accreting  on larger timescales, with respect to our short integration time.
\par
{ The accretion rate reaches a constant maximal value close to the boundary of the zero energy region (see Fig. \ref{Fig:SliceSpherical} and Eq. \ref{energy}). This means that gas accreted by the planet  piles up in the outer envelope in the {\bf N} case, and in the CPD in the other two cases;  in agreement with our dynamical study}. \par
We observe that in the \textbf{nominal} case, the rate of inflow  is of $\sim 0.1 M_{\oplus}/yr$ 
(Table \ref{table:tab3}, in agreement with \cite{LLNCM19}, Fig. 2) 
whilst the accretion rate is of  $ \sim 0.029 M_{\oplus}/yr$. 
This is due to the large exchange of gas in the Hill sphere and corresponds to the qualitative description we  described in the previous section, via the streamlines.\par
Instead, in the {\bf Lm} case, the rate of inflow ($\sim 0.011 M_{\oplus}/yr$, Table \ref{table:tab3})  is marginally higher than
the accretion rate ( $~0.009 M_{\oplus}/yr$, Table \ref{table:tab3}). 
Therefore, although there is a difference of about a factor of 10 in the total inflow rate between the \textbf{nominal} and the {\bf Lm} case, as we expect for a viscous transport proportional to $\Sigma \nu$, there is only a factor of$~3$ in difference for the accretion rate between the two cases.  
This result was not expected, since  the accretion is considered to behave proportionally to the mass inflow rate; however, we have seen that the accretion depends on the dynamics in the Hill sphere and this changes drastically from the nominal case in comparison to the other two cases, which appear to accrete the in flowing gas more efficiently.   
Precisely, the accretion appears to be efficient also in the {\bf LmLv} case,  with an accretion rate of  $~ 0.0015  M_{\oplus}/yr$ for a total inflow rate  of $~ 0.0023  M_{\oplus}/yr$.{ From the viscous disk evolution, the quantity of gas that may be supplied by the disk and can flow onto the planet is proportional to $\Sigma \nu$. The mass inflow on the Hill sphere provided in Table \ref{table:tab3} actually scales as expected based on  viscous evolution}. Instead, for the net accretion rate in the Hill sphere, we have obtained  a scaling  $\propto \sqrt \Sigma \nu$.
\par
 The accretion rate inside the Hill sphere is often used to infer the  mass doubling time of the planet (or of the planetary system when considering the envelope and the circum-planetary disk). In our case, the Jupiter-mass-doubling time is of the order of  a few $ 10^{4}$ of years, as previously reported in the literature,  for both  the {\bf N} case and the low-mass disk case. Therefore, the typical scenario of a Jupiter-mass planet entering runaway accretion late in the lifetime of a protoplanetary disk remains valid when considering disks less massive than  MMSN at the planet's location. \par
In the case of a low-viscosity disk, the mass doubling time increases by an order of magnitude. Thus, it is only in the low-viscosity case that a  giant planet may enter in  runaway accretion earlier in the disk's lifetime to achieve the same final mass.




\section{Conclusion}
\label{Sec:Conclusion}
Using 3D hydrodynamical simulations with a fully radiative equation of state, we have studied the dynamical properties of gas in the Hill sphere of a Jupiter-mass planet in three different protoplanetary disks configurations. The first features a disk with a surface density corresponding to the minimum mass solar nebula (at the planet's location), characterised by constant viscosity, equivalent  to having an alpha parameter of  $4\,10^{-3}$ at the location of the planet.  The second  disk model differs from the nominal one in mass, with a surface density that is ten times smaller than the nominal one and same viscosity; whilst in the third model, we also reduce the viscosity value by a factor of 10.\par
During gap formation, giant planets accrete gas inside the Hill sphere   from the local reservoir. Gas is heated by compression and cools according to opacity, density, and temperature values. This process determines the thermal energy budget inside the Hill sphere. 
We can then describe the characteristics of the planetary system and the dynamics in the Hill sphere when the gas has reached a 
quasi-stationary state.\par
In the nominal case, a pressure-supported envelope forms, as already reported in the literature \citep{2009MNRAS.397..657A, Szulagyietal2016,2019ApJ...887..152F} and gas cools on timescales that are larger than the orbital period on the whole Hill sphere. 
In the other cases, we observe an envelope and a  CPD-like structure supported partially by pressure and partially by rotation.
A necessary condition for the formation of a rotational supported structure was recently given in \cite{2024arXiv240214638K} based upon the requirement that the cooling time scale at the tidal truncation radius must be comparable to (or shorter than) one orbital period at the planet's location. We show that our disk model with low surface density and  and low viscosity  satisfies this criterion, whilst the low surface density case  marginally satisfies it; here, the disk structure in the region 0.2 - 0.4 $R_H$ is indeed quite heavily supported  by pressure.
The dynamics inside the Hill sphere is also shown to depend on the total energy budget, with the gas being accreted preferentially from midplane in the nominal case and lifted up to colder layers, leaving the Hill sphere preferentially at an altitude  of $0.3<z<0.7  R_{H}$. In the other cases, the gas enters in the Hill sphere and accretes on the planetary envelope except for z values very close to the midplane where the radial motion of the gas (relative to the planet) is positive.
The polar accretion channel exists in all the cases although a very small fraction ($~5\%$ in the {\bf N} case and $~1\%$ in the {\bf Lm} and {\bf LmLv} cases) of the accreted gas comes from polar directions with $z>0.8 R_H$.\par
In all the cases, a hot envelope is formed in the planet's vicinity extending to $\sim ~0.4 R_H$, $\sim ~0.2 R_H$ and $\sim ~0.1 R_H$,  respectively, in the 1) nominal, 2) low-mass, and 3) low-mass+low-viscosity cases. A similar result of  giant planets forming an envelope (and 
eventually a disk structure) that is limited to  a fraction of the Hill sphere was also found in \citep{Schuliketal2020,2024arXiv240214638K}. 
 \par
To form a rationally supported structure or CPD for our nominal disk parameters it is necessary to reduce the opacity \citep{Schuliketal2020}. 
In a reduced opacity environment the building blocks for satellite formation may be planetesimals captured in the CPD \citep{2020A&A...633A..93R}.
In this paper, we kept the dust to gas ratio to usual circum-stellar disk values and considering that, although  dust is filtered at the planet's outer gap edge, particles coupled to the gas are free to cross the gap; thus, this may allow the opacity to remain at disk values \citep{2023A&A...675A..75M}.\par
Although we are limited in resolution and a finite smoothing length,  we observe that 
 by increasing the resolution, the inner envelope becomes  hotter and denser; at the same time, the properties in the outer part of the Hill sphere remain basically unchanged, so that our results seem to be robust with respect to resolution and smoothing length (Appendix \ref{sec:AppendixB} and \cite{Szulagyietal2016}).
Concerning the accretion rate and the planet's mass doubling time, our results are consistent with the literature in the nominal and low mass cases. \par
In particular, we have shown that  the scaling of  the net accretion rate  among the
three models behaves as $\sqrt\Sigma \nu$.  Therefore, only 
reducing the mass  does not  significantly increase the mass-doubling time. Low-viscosity disks instead favour a significant increase in the mass-doubling time.
\par

When considering the low-viscosity disk case, we must remark that viscous transport  gives a  star accretion rate of $\sim ~10^{-9} M_{\odot}$ per year, which is  low with respect to observed values of $10^{-7}, 10^{-8} M_{\odot}$ per year \citep{1998ApJ...495..385H,2016A&A...591L...3M}.  Star accretion on surface layers, as provided by  MHD models, offers a mechanism for disks with low  bulk viscosity  to have star accretion rates consistent with  the observed values.
Recently, \cite{2023A&A...670A.113N} measured accretion rates on Jupiter-mass planets by using a simple hydrodynamical model with  star accretion on surface layers induced through angular momentum loss  (with a rate of $\sim ~10^{-8} M_{\odot}$ per year in magnetised winds). 
These authors monitored the gas entering in the Hill sphere of a Jupiter-mass planet 
by varying the thickness 
of the accretion layer and found a  mass-doubling timescale going from 0.2 to 10 Myrs   with larger values for very thin accretion layers.
 The value obtained in this work for the Jupiter doubling time of the planetary system   in the {\bf LmLv} case is of about  0.2 Myrs, consistent with  the case of gas
 radially transported across the whole column density in the \citep{2023A&A...670A.113N}.
Applying the procedure presented here on models  with thin accretion layers is beyond the scope of this paper and  will be undertaken in a future study.\par
 In a follow-up paper (\cite{Cridland2024}, submitted) we explore the chemical evolution of the gas along streamlines and the resulting chemical structure of the gas in the vicinity of the planet. 
The aforementioned dependence of the gas temperature on the spatial resolution could change the chemical structure of the gas in the planet's neighbour and modify the energy content of any molecular line emission that would emanate from near the planet. However, when reaching  temperatures above $2000^\circ K$ by increasing the resolution and further deepening the potential, the model should  take into account hydrogen dissociation, which is not included in this work.{  One example of  an equation-of-state adapted  to take into account the endothermic reaction of hydrogen dissociation was done through 
the reduction of the adiabatic index  by \cite{2019EPSC...13.1408F}. The authors observed temperatures exceeding $10^4 \,^\circ K$ in the few grid cells with very deep gravitational potential  in cases where a CPD is formed.
 This is an indication that performing higher resolution studies with a deeper potential would mainly affect the innermost regions close to the planet. However, we consider that further} numerical experiments with improved models that take into account hydrogen dissociation, as well as observations, are needed. This will help to better constrain the properties of the gas surrounding embedded planets.

\begin{acknowledgements}
{ We thank an anonymous referee for constructive suggestions that helped improving the manuscript.}
MB and AC received funding from the European Research Council (ERC) under the European Union’s Horizon 2020 research and innovation programme (PROTOPLANETS, grant agreement No. 101002188).  AM is grateful for support from the ERC advanced grant HolyEarth N. 101019380. This work was granted access to the HPC resources of IDRIS under the allocation 2023 A0140407233 and 2024 A0160407233 made by GENCI. EL wish to thank Alain Miniussi for maintenance and re-factorisation of the code FARGOCA. 
\end{acknowledgements}
%
%
\bibliographystyle{aa} 
\bibliography{planets}
\begin{appendix} 

\section{Thermal structure}
\label{sec:AppendixA}
Disks that differ in mass and viscosity have also a different vertical thermal structure. 
 To obtain a coherent description of the thermal structure, we included stellar irradiation in our models using the numerical prescription provided in \cite{Bitschetal2014} and performed 2D $(r,z)$ 
 axisymmetric simulations. We considered  a solar-mass star with a surface temperature of $4370 ^\circ K$, a radius of 1.5 solar radii, and  a disk annulus that extends in the interval $r/r_p=[0.4,9]$ (or $r=[2.08,45.9] $ au).  In the vertical direction  the disk extends by $20^\circ$ over the midplane. We considered a resolution of $(N_r,N_\theta)=(528,82)$. \par
In our 3D simulations  we did not consider stellar irradiation (Eq. \ref{eq:Edot2Temp}), in order to have reasonable computational times and to limit the amount of CPUs. However,
we can mimic the thermal equilibrium structure obtained in disks heated by the star, by suitably fixing the disk surface temperature. \par
To this purpose, we ran  test cases in which we imposed different values of the boundary temperature, $T_b,$ at the disk surface. The simulations have a resolution of $(N_r,N_\theta)=(228,32)$ on a disk annulus extending radially in the interval $r/r_p=[0.4,2.5]$ and vertically by about  $7 ^\circ$ over the midplane \footnote{i.e. the typical disk domain for 3D simulations with embedded planets in the  paper}.
\par
\begin{figure*}
\centering
\includegraphics[width=0.8\textwidth]
   {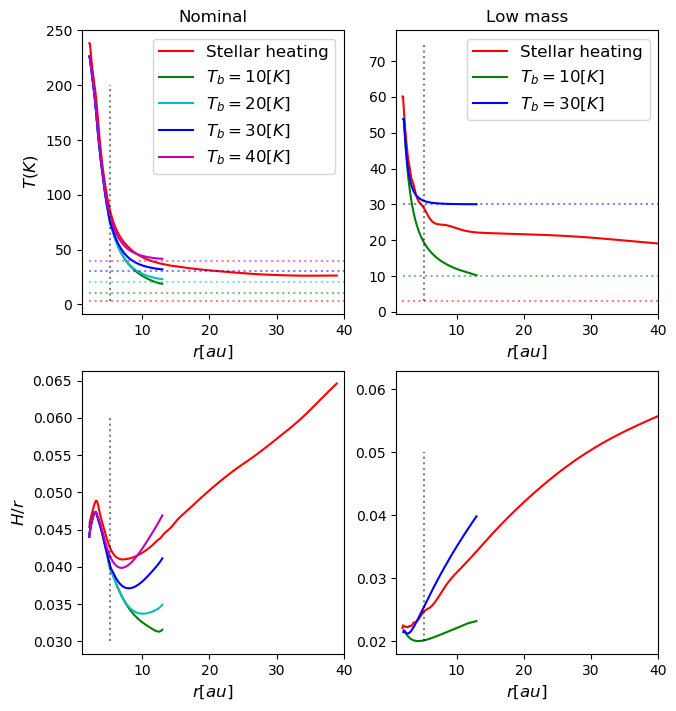}
      \caption{Midplane temperature ({\bf top}) and disks' aspect ratio ({\bf bottom}) as a function of radial distance from the host star for 2D simulations for the nominal and {\bf Lm} cases (left and right, respectively). The red lines include the effect of stellar irradiation. The other simulations have no stellar heating and differ in the boundary
      temperature, $T_b$, at  disk's surface.  The value of $T_b$ that allows us to recover the profile of the stellar irradiated disk is used for 3D simulations of
      in the rest of the paper. Horizontal dotted lines in the top plots correspond to the values of $T_b$ in the simulations. The vertical black dotted line marks  the position of a  Jupiter-mass planet in the paper. %
      }
         \label{Fig:Thr} 
\end{figure*}

Figure \ref{Fig:Thr}, left shows the equilibrium midplane temperature (top)
and the disk aspect ratios (bottom) for the nominal {\bf N} model including the stellar heating and for the four test cases
without stellar heating.
In the inner part of the disk, the values of the aspect ratio are very similar for disks with and  without stellar heating (see also \citealt{Bitschetal2013}).
This means that viscous heating dominates over  the stellar irradiation up to $7-8$ au. Beyond 10 au, the  irradiated disk is flared, while the disk with $T_b = 10 $ $^\circ K$ collapses to small $H/r$ at large radii, as is typical of disks without stellar heating. We mimicked the flaring by imposing a temperature at the disk's surface higher than $T_b = 10 $ $^\circ K$. For the nominal model,  a value of $T_b = 40$ $^\circ K$ nicely fits the temperature structure of the stellar irradiated disk. We remark  that the planet position in the paper (vertical black dotted line in Fig.{Fig:Thr}) is 
at the boundary of the region dominated by viscous heating.
\par
When considering the {\bf Lm} disk  (Fig.\ref{Fig:Thr}, right), the viscous heating, proportional to $\Sigma$, dominates only in a very narrow region at the inner disk boundary (Fig. \ref{Fig:Thr}, bottom right), and the planet's location is  in the part of the disk where stellar heating dominates. 
The boundary temperature required to mimic stellar irradiation is $T_b=30$ $^\circ K$.
In the model with low viscosity ({\bf LmLv }), the stellar heating dominates in the full disk range that we are considering. We remark  that the amount of energy absorbed in the disk from the star's photons depends on the density and the opacity and therefore the {\bf Lm} and {\bf LmLv} disks have the same thermal structure \footnote{except for the small inner bump in the {\bf Lm} disk where viscous heating dominates; see Fig,\ref{Fig:Thr}, bottom right}. Therefore, we also considered a value of $T_b$  of $30$ $^\circ K$  in the low-viscosity case.

\section{Dependence of results on resolution and smoothing length}
\label{sec:AppendixB}
\begin{figure*}
\centering
\includegraphics[width=\textwidth]{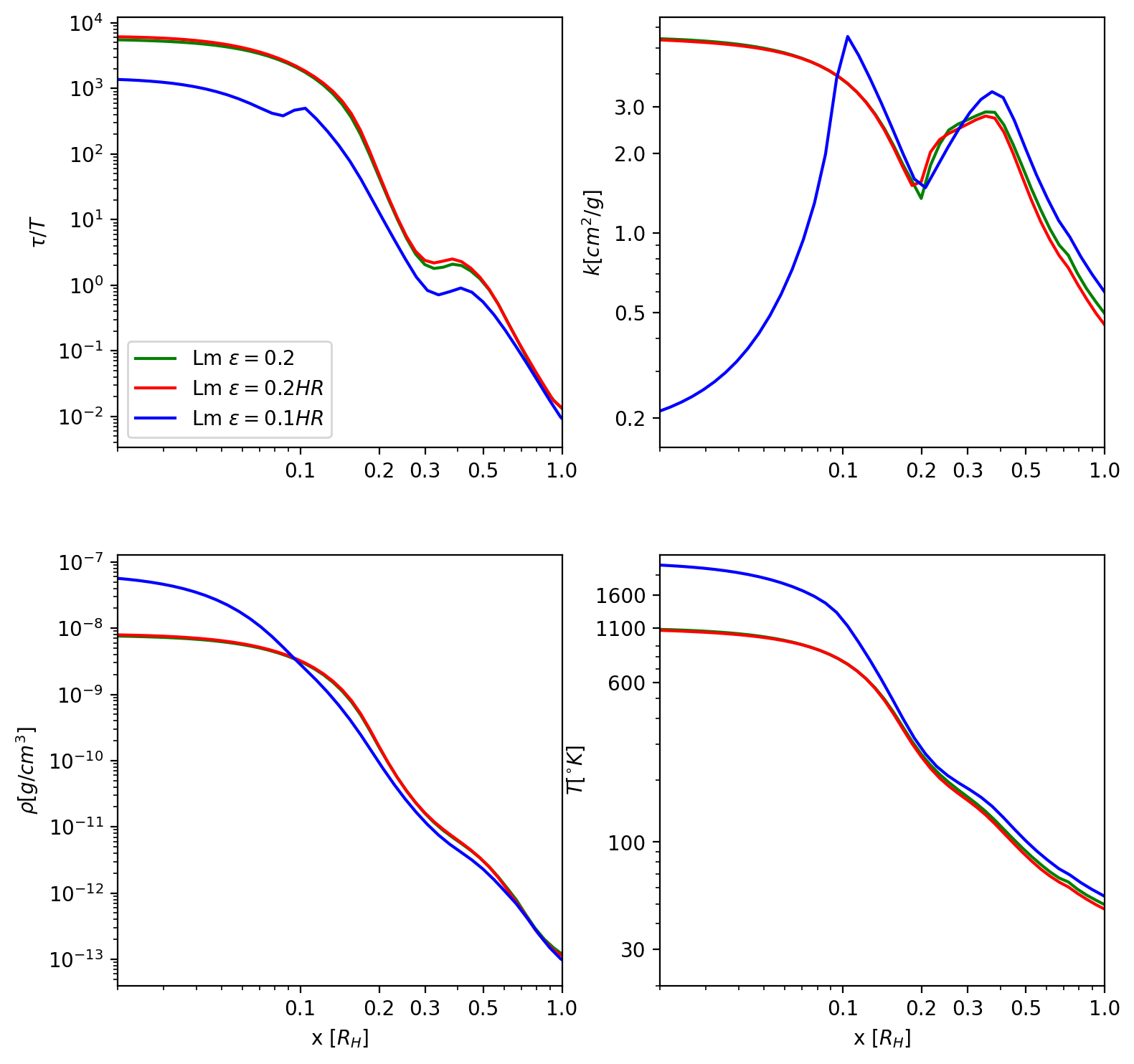} 
   \caption { Same as Fig. \ref{Fig:Shells} but for the {\bf Lm} 
   case {(green curves) and for higher resolution simulations with: a smoothing length  $\epsilon = 0.2$ {(red curves)  and $\epsilon = 0.1$ (blue curves)}}. We remark that  decreasing the smoothing length impacts the innermost regions ($x <0.1 R_{H}$) leaving unchanged the outer part of the Hill sphere.}
\label{Fig:ShellsHR} 
\end{figure*}

This appendix presents a study of the dynamic of the gas in the planet's vicinity for the low mass  case  by increasing the resolution and by decreasing the smoothing length in order to check the convergence of results with respect to both quantities: the resolution and the smoothing length. 
For this, we further refined our grid on the same domain previously studied (phase C)  and restarted the code from the output at five orbital periods of phase C for two additional orbits on a grid of  $(N_r,N_\varphi,N_\theta)=(498,1998,60)$ grid-cells
 for two  cases, shown in Fig. \ref{Fig:ShellsHR}: \begin{itemize}
\item { We kept a smoothing length of $\epsilon = 0.2$. Results  (red curves in Fig.\ref{Fig:ShellsHR}) are shown over-plotted with respect to those of the {\bf Lm} case (presented in this paper) at the resolution of phase C (green curves) and showing convergence with respect to resolution.}
\item 
{ The new grid has ten grid cells inside 0.1 Hill radii.\ Therefore, according to the criterion from \cite{LLNCM19}, by requiring eight grid cells in the smoothing length} we can set $\epsilon $ to  $0.1$ $R_H$. On the first { restarted} orbit we smoothly reduced the smoothing length and we show results (blue curves in Fig. \ref{Fig:ShellsHR}) at the end of the second orbit when  quasi-stationary fields are obtained.
\end{itemize}

 The nominal case was analysed in \cite{Szulagyietal2016} using nested meshed with the Jupiter's code and by simulating disks sectors  with the code fargOCA. The resulting gas structure around the planet showed a spherical envelope, like the one that we obtain here at lower resolution, with temperatures increasing in the innermost grid-cells when increasing the resolution and 
decreasing the smoothing length, leaving almost unchanged the density and temperature in the outer part of the Hill sphere.
The same conclusion can be drawn here: we notice that the temperature increase inside 0.1 $R_H$ with a corresponding decrease in the opacity, which results in a decrease in the radiative timescale. The planetary envelope results to be denser but the profiles for distances larger than 0.1 $R_H$ remain unchanged. Therefore, we conclude that our results are robust with respect to both the resolution and the  smoothing  length.



\end{appendix}

\end{document}